\documentclass[pre,aps,amssymb,amsmath,superscriptaddress,twocolumn]{revtex4}

\usepackage{graphicx}
\usepackage{dcolumn}
\usepackage{bm}
\usepackage{epstopdf}

\newcount\driver
\newcount\bozza

scaled\magstep1                  
scaled\magstep1 
 
scaled\magstep1                 \font\indbf=cmbx10 scaled\magstep2

scaled \magstep2

{\count255=\time\divide\count255 by 60
\xdef\hourmin{\number\count255}
        \multiply\count255 by-60\advance\count255 by\time
   \xdef\hourmin{\hourmin:\ifnum\count255<10 0\fi\the\count255}}

\let\a=\alpha \let\b=\beta    \let\g=\gamma     \let\d=\delta     \let\e=\varepsilon
  \let\h=\eta           \let\l=\lambda
\let\m=\mu    \let\n=\nu      \let\x=\xi                \let\r=\rho
\let\s=\sigma \let\t=\tau            
\let\ps=\psi   \let\o=\omega     
 \let\D=\Delta

\def\VV{{\cal V}}
\def\WW{{\cal W}}
\def\TT{{\cal T}}
\def\RR{{\cal R}}\def\LL{{\cal L}}

\def\xx{{\bf x}}
  
\def\yy{{\bf y}}\def\kk{{\bf k}}\def\nn{{\bf n}}

       \def\oo{{\underline \omega}}
\def\ee{{\underline \varepsilon}}

\let\==\equiv

\let\io=\infty
\let\0=\noindent

\def\*{{\hfill\break\null\hfill\break}}

\def\tilde#1{{\widetilde #1}}

\def\aps{{\it a posteriori}}

\def\tende#1{\,\vtop{\ialign{##\crcr\rightarrowfill\crcr
             \noalign{\kern-1pt\nointerlineskip}
             \hskip3.pt${\scriptstyle #1}$\hskip3.pt\crcr}}\,}
\def\otto{\,{\kern-1.truept\leftarrow\kern-5.truept\to\kern-1.truept}\,}

\def\wh#1{\widehat{#1}}
\def\hat#1{\wh{#1}}
\def\sqt[#1]#2{\root #1\of {#2}}

\def\bp{{\bar \ps}}

\def\VV{{\cal V}}
\def\WW{{\cal W}}
\def\TT{{\cal T}}
\def\RR{{\cal R}}\def\LL{{\cal L}}

\def\T#1{{#1_{\kern-3pt\lower7pt\hbox{$\widetilde{}$}}\kern3pt}}
\def\VVV#1{{\underline #1}_{\kern-3pt
\lower7pt\hbox{$\widetilde{}$}}\kern3pt\,}
\def\W#1{#1_{\kern-3pt\lower7.5pt\hbox{$\widetilde{}$}}\kern2pt\,}

\def\indica{\leaders \hbox to 0.5cm{\hss.\hss}\hfill}
\def\guida{\leaders\hbox to 1em{\hss.\hss}\hfill}
\mathchardef\oo= "0521

\def\xx{{\bf x}}
\def\yy{{\bf y}}\def\kk{{\bf k}}\def\nn{{\bf n}}

\def\oo{{\underline \omega}}

\def\qed{\raise1pt\hbox{\vrule height5pt width5pt depth0pt}}
 
  \def\bp{{\bar p}} 

\def\indic{\hbox{\raise-2pt \hbox{\indbf 1}}}

\def\ins#1#2#3{\vbox to0pt{\kern-#2 \hbox{\kern#1 #3}\vss}\nointerlineskip}

\newdimen\xshift \newdimen\xwidth \newdimen\yshift

\def\insertplot#1#2#3#4#5#6{%
\xwidth=#1pt \xshift=\hsize \advance\xshift by-\xwidth \divide\xshift by 2%
\begin{figure}[ht]
\vspace{#2pt} \hspace{\xshift}
\begin{minipage}{#1pt}
#3 \ifnum\driver=1 \griglia=#6
\ifnum\griglia=1 \openout13=griglia.ps \write13{gsave .2
setlinewidth} \write13{0 10 #1 {dup 0 moveto #2 lineto } for}
\write13{0 10 #2 {dup 0 exch moveto #1 exch lineto } for}
\write13{stroke} \write13{.5 setlinewidth} \write13{0 50 #1 {dup 0
moveto #2 lineto } for} \write13{0 50 #2 {dup 0 exch moveto #1
exch lineto } for} \write13{stroke grestore} \closeout13
\includegraphics{griglia.ps} \fi
\includegraphics{#4.ps}\fi%
\ifnum\driver=2 \fi
\end{minipage}
\caption{#5}
\end{figure}
}

\newdimen\shift \shift=-1.5truecm
\def\lb#1{%
\ifnum\bozza=1
\label{#1}\rlap{\hbox{\hskip\shift$\scriptstyle#1$}}
\else\label{#1} \fi}

\def\be{\begin{equation}}
\def\ee{\end{equation}}
\def\bea{\begin{eqnarray}}\def\eea{\end{eqnarray}}
\def\bean{\begin{eqnarray*}}\def\eean{\end{eqnarray*}}
\def\bfr{\begin{flushright}}\def\efr{\end{flushright}}
\def\bc{\begin{center}}\def\ec{\end{center}}
\def\bal{\begin{align}}\def\eal{\end{align}}
\def\ba#1{\begin{array}{#1}} \def\ea{\end{array}}
\def\bd{\begin{description}}\def\ed{\end{description}}
\def\bv{\begin{verbatim}}\def\ev{\end{verbatim}}
\def\nn{\nonumber}
\def\Halmos{\hfill\vrule height10pt width4pt depth2pt \par\hbox to \hsize{}}
\def\pref#1{(\ref{#1})}

\def\ins#1#2#3{\vbox to0pt{\kern-#2 \hbox{\kern#1 #3}\vss}\nointerlineskip}

\newdimen\xshift \newdimen\xwidth \newdimen\yshift
\newcount\griglia

\def\insertplot#1#2#3#4#5#6{%
\xwidth=#1pt \xshift=\hsize \advance\xshift by-\xwidth \divide\xshift by 2%
\begin{figure}[ht]
\vspace{#2pt} \hspace{\xshift}
\begin{minipage}{#1pt}
#3 \ifnum\driver=1 \griglia=#6
\ifnum\griglia=1 \openout13=griglia.ps \write13{gsave .2
setlinewidth} \write13{0 10 #1 {dup 0 moveto #2 lineto } for}
\write13{0 10 #2 {dup 0 exch moveto #1 exch lineto } for}
\write13{stroke} \write13{.5 setlinewidth} \write13{0 50 #1 {dup 0
moveto #2 lineto } for} \write13{0 50 #2 {dup 0 exch moveto #1
exch lineto } for} \write13{stroke grestore} \closeout13
\includegraphics{griglia.ps} \fi
\includegraphics{#4.ps}\fi%
\ifnum\driver=2 \fi
\end{minipage}
\caption{#5}
\end{figure}
}

\newdimen\shift \shift=-1.5truecm
\def\lb#1{%
\label{#1}\rlap{\hbox{\hskip\shift$\scriptstyle#1$}}
\else\label{#1} \fi}

\def\be{\begin{equation}}
\def\ee{\end{equation}}
\def\bea{\begin{eqnarray}}\def\eea{\end{eqnarray}}
\def\bean{\begin{eqnarray*}}\def\eean{\end{eqnarray*}}
\def\bfr{\begin{flushright}}\def\efr{\end{flushright}}
\def\bc{\begin{center}}\def\ec{\end{center}}
\def\bal{\begin{align}}\def\eal{\end{align}}
\def\ba#1{\begin{array}{#1}} \def\ea{\end{array}}
\def\bd{\begin{description}}\def\ed{\end{description}}
\def\bv{\begin{verbatim}}\def\ev{\end{verbatim}}
\def\nn{\nonumber}
\def\Halmos{\hfill\vrule height10pt width4pt depth2pt \par\hbox to \hsize{}}
\def\pref#1{(\ref{#1})}

\usepackage{amsmath}
\usepackage{amsfonts}
\usepackage{amssymb}
\usepackage{epstopdf}

scaled\magstep1

\let\a=\alpha \let\b=\beta  \let\g=\gamma  \let\d=\delta
\let\e=\varepsilon
  \let\h=\eta     \let\l=\lambda
\let\m=\mu    \let\n=\nu    \let\x=\xi         \let\r=\rho
\let\s=\sigma \let\t=\tau    
\let\ps=\Psi   \let\o=\omega
 \let\D=\Delta

 \def\VV{{\cal V}}
 \def\WW{{\cal W}}
\def\TT{{\cal T}} 
\def\RR{{\cal R}}\def\LL{{\cal L}}

 \def\xx{{\bf x}} \def\yy{{\bf y}} 
\def\kk{{\bf k}}

\def\nn{\nonumber}

\def\\{\hfill\break}
\def\={:=}
\let\io=\infty
\let\0=\noindent

\def\tende#1{\,\vtop{\ialign{##\crcr\rightarrowfill\crcr\noalign{\kern-1pt
    \nointerlineskip} \hskip3.pt${\scriptstyle #1}$\hskip3.pt\crcr}}\,}
\def\otto{\,{\kern-1.truept\leftarrow\kern-5.truept\to\kern-1.truept}\,}

\def\wh{\widehat}
\def\to{\rightarrow}

\def\qed{\hfill\raise1pt\hbox{\vrule height5pt width5pt depth0pt}}

\def\be{\begin{equation}}
\def\ee{\end{equation}}
\def\bp{\begin{pmatrix}}
\def\ep{\end{pmatrix}}
\def\bea{\begin{eqnarray}}
\def\eea{\end{eqnarray}}
\def\nn{\nonumber}
\def\pref#1{(\ref{#1})}

\def\lb{\label}

\begin{document}

\title{Dense gaps and scaling relations in the interacting Aubry-Andr\'e model}
\author{Vieri Mastropietro}
\affiliation{Universit\`a degli Studi di Milano, 
Via Saldini, 50, 20133 Milano - Italy}

\begin{abstract} 
We study, by rigorous Renormalization Group methods,
the interacting Aubry-Andr\'e model 
for fermions in the extended regime.
We show that
the infinitely many gaps of the single particle
spectrum persist in presence of weak many body interactions,
despite the presence of Umklapp large momentum processes
connecting the Fermi points. The width of the gaps
is strongly renormalized through critical exponents which verify
exact scaling relations. 
\end{abstract}


\maketitle


\section{Introduction}

The interplay between disorder and many body interaction produces intricate phenomena whose theoretical understanding is extremely difficult, especially in high dimensions.
It is natural then to consider one dimensional models and in this context a major role is played by
the interacting Aubry-Andr\'e model \cite{AA},\cite{AA1}, in which a quasi-random disorder in a one dimensional lattice is
represented by a quasi-periodic potential  $u\cos (2\o\pi x+\phi)$, with $\o$ irrational.
This choice of disorder
is realized in cold atoms experiments; using two  
incommensurate optical lattices one can simulate the Aubry-Andr\'e model with bosons
\cite{I1},\cite{I3} or fermions \cite{I2}, and theoretical predictions can be compared with experimental results for a wide range of parameters.
The interacting Aubry-Andr\'e model catches a number of properties 
of more realistic systems; for instance 
it has a metal-insulator transition 
at a single particle level as in 3d random models.
In addition, it exhibits non trivial topological 
properties \cite{t1}. While the properties of the non interacting Aubry-Andr\'e model 
are well understood, 
several basic properties of the interacting case are still not known.

We consider the fermionic interacting Aubry-Andr\'e model with hopping $t$ and two-body interaction with strength $U$, which
can be mapped in Heisenberg chains with a quasi-periodic magnetic field, and in one dimensional bosons with hard-core interaction. In the absence of quasi random disorder $u=0, U\not=0$ the many body interaction produces  {\it Luttinger liquid behavior} \cite{Ha} characterized by anomalous critical exponents
verifying exact scaling relation.
In the opposite limit $U=0,u\not=0$ the fermions are only subject to the disorder and their behavior
can  be deduced by the corresponding single particle Schroedinger equation
({\it almost-Mathieu} equation), which has extended eigenstates for $u<2t$ (metallic regime) \cite{Si1}\cite{E}
and localized eigenstates for $u>2t$ (Anderson insulator) \cite{FSW}.
In addition the 
spectrum is a Cantor set for any irrational $\o$ \cite{A1}.
Such remarkable properties are related to a deep connection 
of the properties of the almost Mathieu equation with 
the {\it Kolmogorov-Arnold-Moser} (KAM) theorem expressing the stability of invariant tori in quasi integrable hamiltonian systems. 
In the extended regime $u<2t$, there are infinitely many gaps of decreasing size, forming a dense set;
the zero temperature correlations of 
the fermionic many body non interacting problem
decay exponentially 
if the chemical potential is in one of the gaps, while as a power law for chemical potentials in a point of the spectrum \cite{BGM}.

Given the complexity of the single particle problem, and the critical behavior produced by the 
many body interaction on free particles,
it is not surprising that
quantitative predictions for the interacting Aubry-Andr\'e model  
are a difficult task, and have been pursued mostly by numerical methods
in the fermionic  \cite{Ri}, \cite{H4}, \cite{H5},\cite{ss} and bosonic case
 \cite{S},\cite{T},\cite{G}; non perturbative effects could however be missed due to size limitations.
On the analytical side, in \cite{M1} it was studied a system of interacting fermions subject to a weak quasi-periodic potential with exponentially fast decaying harmonics, and insulating behavior was rigorously proved for chemical potentials in the gaps, provided that the corresponding harmonic is non vanishing; for 
Aubry-Andr\'e potential, this says that the largest gap persists.
In \cite{G2},\cite{G3} it was studied a quasi periodic potential with slow 
decaying harmonics, the  
Fibonacci potential \cite{SM}, and second order analysis suggests a
a scenario (whose validity at all orders
has to be checked) in which
the largest gaps persist in presence of interaction while the smaller ones 
are closed by an arbitrarily small interaction, so that an insulator-metal transition is present at weak coupling.

In the present paper we consider 
the fermionic interacting  
Aubry-Andr\'e model in the extended  regime (for the localized regime see \cite{M2}) and
we address the following questions: do small gaps close or they all persists
in presence of a weak interaction? If the system remains insulating, which is the interaction effect on the gaps?   Due to the incommensurablity of the potential,
the opposite Fermi points can be (almost) connected by Umklapp processes involving the exchange of very high momenta, producing a sort of infrared-ultraviolet mixing problem;  
the persistence or closure of small gaps 
is therefore a subtle problem related to the relevance or irrelevance of such processes, 
and it requires non-perturbative methods and the exploitation of number theoretical properties
to be settled. Our conclusion, based on rigorous Renormalization Group methods methods,  is that 
every gap 
persists 
in presence of a weak many body interaction; in contrast to what is expected for Fibonacci potential, no insulator-metal transition is present.
The width of the gaps $\D_{n,U}$, $n=1,2,,,$, obtained by the decay rate of correlations, 
is strongly renormalized by the interaction through a critical exponent
$\D_{n,U}\sim \D_{n,0}^{X_n}$, with $X_n$ verifying the exact {\it scaling relation}
$X_n={1\over 2-K_n}$, with $K_n$
is the exponent of the oscillating part of the density-density correlation; note that $K_n$
depends weakly from $n$ and is greater or smaller than $1$ depending on negative or positive $U$.
The fact that in the interacting Aubry-Andr\'e
there is a natural scale 
with such exponent
is apparently observed in numerical simulations \cite{S}.
As this renormalization is present in all gaps,
including the smallest ones, the relative size of renormalized and bare gaps is dramatically increased or decreased depending on the repulsive or attractive nature of the interaction. 

The paper is organized in the following way. In \S II we define the model, and we discuss some number theoretical properties of the frequency; in \S III we
set up our exact Renormalization Group analysis.
According to dimensional considerations, there is an infinite number of relevant or marginal interactions;
however, in \S IV we will show that indeed a huge number of such terms is indeed irrelevant
and only a finite number of effective interactions is relevant or marginal. 
In \S V we renormalize the true relevant or marginal terms,
and in \S VI we study the flow of the running coupling constants. Finally in \S VII we prove the persistence of the gaps and the validity of the scaling relations and in \S VIII the main conclusions are presented.

\section{The model}

The Hamiltonian of the fermionic interacting Aubry-Andr\'e model, as appearing in cold atoms experiments, 
see {\it e.g.}
\cite{I2}, is ($t=1$ for definiteness)
\bea &&H=-\sum_x {1\over 2}(a^+_{x+ 1} a_{x}+ a^+_{x} a^-_{x-1} )+\m \sum_x 
 a^+_{x} a^-_{x}\\
&&+u\sum_x  \cos (2 \pi \o x+\phi)
a^+_{x} a^-_{x}+U \sum_{x,y}v(x-y)
a^+_{x} a^-_x a^-_{y} a^+_{y}
\label{1.1}\nn\eea
with $x=0,\pm 1,\pm 2,...$, $u>0$, $v(x-y)$ a short range interaction and
$a^\pm_x$ fermionic creation or annihilation operators.
It is not restrictive to choose the phase $\phi=0$. 

The irrational frequency $\o$ is assumed {\it Diophantine}
\be ||2 n \pi\o||\ge C
|n|^{-\t},\quad\quad n\not=0\label{d}\ee 
$||.||$ being the norm on the one dimensional $2\pi$ torus. This is 
the standard condition usually assumed for the non interacting case $U=0$
\cite{Si1},\cite{E},\cite{FSW} and 
it is physically not restrictive as Diophantine numbers
have full measure. 

In order to impose periodic boundary conditions (usually adopted in 
numerical simulations ) we recall that
the continued fraction representation  of a number $\o$
is the sequence of positive integers such that
\be
\o=a_0+{1\over a_1+{1\over a_2+{1\over a_3+...}}}
\ee
If $\o$ is rational the sequence $a_0,a_1,.,,,a_N$ is finite, while if $\o$ is irrational 
the sequence is infinite. In noble numbers, which are Diophantine, the sequence has the form $a_0;a_1,..,a_{N},1,1,...1,..$. For instance
the {\it golden ratio} $\o={\sqrt{5}+1\over 2}$ has representation $1;1,..1,..$ and
it verifies the Diophantine condition \pref{d} with $\t=1$ and $C_0={3+\sqrt{5}\over 2}$.
Periodic boundary conditions are imposed considering
a sequence of periodic potentials of period $L$ such that 
quasi-periodicity is recovered in the thermodynamic limit.
In order to do that 
we approximate
$\o$ by a sequence of rational numbers ({\it convergents}) 
${p_1\over q_1}=a_0+{1\over a_1}$, ${p_2\over q_2}=a_0+{1\over a_1+{1\over a_2}}$ and so on. For the golden ratio, the sequence
is given by the ratio of Fibonacci numbers $\{1,{2,\over 1},{3\over 2},{8\over 5},{13\over 8},.,{p_i\over q_i},...\}$. Properties of the convergents imply, see {\it e.g.} \cite{BGM},
that if $\o$ verifies the Diophantine condition
then $|\pi(n{p_i\over q_i}-k)|\ge {C\over 2 |n|^\t}$ if $q_1\le n\le {q_i\over 2}$ and any $k$.
Therefore we can impose periodic boundary conditions by considering a sequence of 
lengths $L_i$ and periodic potentials with
 rational frequencies $\o_{L_i}$, $i=1,2,3,..$ converging to a Diophantine number $\o$;  we choose
$\o_{L_i}={p_i\over q_i}$ and $L_i=q_i$,  so that the analogue of the Diophantine condition 
is verified at finite volume.

We are interested in the 
thermodynamical correlations at zero temperature, like  the 2-point function with imaginary time
$<a^-_\xx a^+_\yy>$, $a_\xx^\pm=e^{H x_0} a^\pm_x e^{-H x_0}$,
$\xx=(x_0,x)$
and 
$<>={{\rm Tr} e^{-\b H} \TT .\over{ \rm Tr} e^{-\b H}}$ and $\TT$ is the time order product. Another important quantity is the density-density correlation
$<\r_\xx;\r_\yy>_T$, with $\r_{\xx}=a^+_{\xx} a^-_{\xx}$  and $T$ denotes truncation.
In the non interacting $U=u=0$ limit
$<\TT a^-_\xx a^+_\yy>|_{U=u=0}=g(\xx,\yy)$ with
\be
g(\xx,\yy)=\int d\kk 
{e^{i\kk(\xx-\yy)} 
\over -i k_0+\cos k -\m}
\label{prop}
\ee
We call $p_F$ the Fermi momentum
defined as  $\m=\cos p_F$; the denominator of \pref{prop}
is vanishing in correspondence of the two Fermi momenta $\pm p_F$.

We will study the system considering both the quasi-periodic
potential and the interaction as perturbations of the free Fermi gas.
It could seem more natural to consider as unperturbed system 
the fermions described by the almost Mathieu equation, whose properties are known.
This however would not work, as the presence of anomalous critical exponents related to the Luttinger
liquid behavior says that an expansion in power series of $U$  around zero cannot be convergent (unless $U$ is much smaller than $u$).

\section{Renormalization Group analysis}

The Fermi momenta in correspondence of which there are gaps are
\be
p_F=n_F \pi \o \quad {\rm mod.} 2\pi\label{pf}
\ee
with $n_F$ integer.
In order to take into account the shift of the chemical potential due to the interaction, we choose a chemical potential of the form $\cos(n_F\pi\o)+\n$, and we will choose
the counterterm $\n=\n(u,U)$ so that the Fermi momentum is just \pref{pf}. 

The correlations can be obtained by derivative of the {\it generating function}, expressed by the following 
Grassmann integral 
\be
e^{W(\phi,J)}=\int P(d\psi)e^{\VV(\psi)+B(\phi,J)}\label{ww}
\ee
where $\psi^\pm_\xx$ are Grassmann variables, $P(d\psi)$ is the Grassmann gaussian integration with propagator
\pref{prop}, with $\m=\cos n_F \pi \o$, $\VV$ is the effective interaction
\bea
&&\VV=-U\int d\xx d\yy v(\xx-\yy)\psi^+_{\xx}\psi^-_{\xx} \psi^+_{\yy}\psi^-_{\yy}+\\
&&u\int d\xx \cos (2\pi\o x)
\psi^+_{\xx} \psi^-_{\xx}+\n \int d\xx
\psi^+_{\xx} \psi^-_{\xx}\nonumber
\eea
with $\int d\xx=\sum_x\int_{-\b/2}^{\b/2}dx_0$, and $B$ is the source term
\be
B(\phi,J)=\int d\xx [\phi^+_\xx\psi^-_{\xx}+\phi^-_\xx\psi^+_{\xx}+J_\xx\psi^+_\xx\psi^-_\xx]
\ee
The 2-point function is given by ${\partial^2 W\over \partial\phi^-_\xx\partial\phi^+_\yy}|_0$
and the density-density correlation is given by 
${\partial^2 W\over\partial J_\xx \partial J_\yy}|_0$. 

The generating functional \pref{ww}
can be analyzed by exact Wilsonian Renormalization Group.
We introduce a smooth cut-off function $\chi_\r(\kk)$, $\kk=(k_0,k)$,which is non vanishing for
$\sqrt{k_0^2+((k-\r p_F)_{\rm mod. 2\pi})^2)}\le \g$, where $\r=\pm 1$ and $\g>1$
is a suitable constant; therefore we can write the propagator as 
\be
\hat g(\kk)=\hat g^{(u.v.)}(\kk)+\sum_{\r=\pm}\hat g_\r(\kk)
\ee where $\hat g_\r(\kk)={\chi_\r(\kk)\over -i k_0+\cos k-\cos p_F}$, 
and correspondingly
 $\psi_{\kk}=\psi^{(u.v.)}_{\kk}+\sum_{\r=\pm 1} \psi_{\kk',\r}$ with $k=k'+\r p_F$, $\kk'=(k_0,k')$.
This simply says that we can write 
the fermionic field as sum of two independent fields living close to one of the Fermi points, up to a regular field. 
We can further decompose \be \hat g_\r(\kk)=\sum_{h=-\io}^0 
\hat g^{(h)}_{\r}( \kk)\ee with 
$\hat g^{(h)}_\r(\kk)$ similar to $\hat g_{\r}(\kk)$ with $\chi$ replaced by $f_h$ 
with,
where $f_h(\kk)$ is non vanishing in a region $\g^{h-1}\le\sqrt{k_0^2+
v_F^2 {k'}^2}\le \g^{h+1}$, with $v_F=\sin p_F$. 
After the integration of $\psi^{(u.v.)}, \psi^{(0)},..,\psi^{(h+1)}$ the generating function has the form
\be
e^{W(\phi,J)}=\int P(d\psi^{(\le h)})e^{\VV^{(h)}(\psi)+B^{(h)}(\psi,\phi,J)}\label{hh}
\ee
where $P(d\psi^{(\le h)})$ has propagator $g_\r^{(\le h)}=\sum_{k=-\io}^h g_\r^{(k)}$ and $\VV^{(h)}(\psi)=$
\be \sum_{m,n,\underline\r} \int d\kk'_1...d\kk'_m  W_{m,n}^{(h)}(\underline \kk')\psi^{\e_1(\le h)}_{\r_1,\kk'_1}...\psi^{\e_m(\le h)}_{\r_m,\kk'_m}\d_{n,m}(\underline\kk')
\label{ep}\ee
where $\d_{n,m}(\underline\kk')$ is a $L\b$ times a periodic Kronecker delta non vanishing for
\be
\sum_{i=1}^{m}\e_i \r_i k'_i=-\sum_{i=1}^{m}\e_i \r_i p_F
+2 n \pi \o+2l \pi \label{hh1}
\ee
with $n=0,1,2$ and $l=0,\pm 1,..$
The kernels $W_{m,n}^{(h)}$ are sum of Feynman diagrams obtained connecting vertices $u$, $U$
or $\n$ with propagators $g^{(k)}$ with $k>h$ ; $B^{(h)}$ is given by a similar expression with the only difference that some of the external lines are associated to $\phi$ or $J$ external fields.
In each of the Feynman diagrams contributing to $W_{m,n}^{(h)}$ there are a set of vertices $u e^{i \s_i \pi\o x}$, $\s_i=\pm$ and 
$n=\sum_i \s_i$. The relation \pref{hh1} is the momentum conservation; when the r.h.s. is vanishing
the momentum {\it measured from the Fermi points} is also conserved.
The single scale propagator has the following form
\be
g_{\r}^{(h)}(\kk')=g_{rel,\r}^{(h)}(\kk')
+r^{(h)}(\kk')\label{rel}
\ee
where $g_{rel,\r}^{(h)}(\kk')=
{f_h(\kk')\over -i k_0+\r v_F k'}$ is the dominant part of the propagator and
$r^{(h)}(\kk')$; therefore in the above RG procedure 
naturally emerges a description in terms of massless relativistic chiral fermions with propagator 
$g_{rel,\r}^{(h)}(\kk')$.  Note also that in the effective potential $\VV^{(h)} $ appear terms with any $n,m$,
and only a few of them were originally present in the initial potential $\VV$.

According to power counting arguments, the quartic terms are {\it marginal} and the quadratic terms are {\it relevant}; all other terms are irrelevant. There are then apparently {\it infinitely} many dimensionally relevant or marginal
terms, depending on the value of $n$ in \pref{ep}. 
%
%
A natural distinction is if the r.h.s. of \pref{hh1} is vanishing or not.
The first case corresponds to processes {\it exactly connecting} the Fermi points;
this happens in the following cases, if $m=2,4$:
a)$m=4$, $n=0$ and $\sum_i \r_i \e_i=0$ (effective interaction
of the form $\psi^+_{+}\psi^-_{+}\psi^+_{-}\psi^-_{-}$);
b)$m=4$, $|n|=4n_F$ and $|\sum_i \e_i \r_i|= 4$ (effective interaction  $\psi^+_{+}\psi^-_{-}\psi^+_{+}\psi^-_{-}$ whose local part is vanishing, so is indeed irrelevant);c) $m=2$, $|n|=0$, $\r_1=\r_2$ (chemical potential $\psi^+_\r\psi^-_\r$); d)
$m=2$, $|n|=2n_F$, $\r_1=-\r_2$ (gap $\psi^+_\r\psi^-_{-\r})$.

The other case is when the r.h.s. of \pref{hh1} in non vanishing
and here comes the main difference between the periodic and quasi-periodic case. 
In the periodic case when $\o$ is rational
the r.h.s. of \pref{hh1} is {\it large} (if non vanishing);
as the fields $\psi^{(\le h)}$ carry a momentum $k'$ with size smaller that $\g^{h}$,
the condition \pref{hh1} cannot be satisfied and such terms are
vanishing for large $|h|$; 
therefore the terms 
with $m=2,4$ such that  r.h.s. of \pref{hh1} is non vanishing are indeed trivially irrelevant in the periodic case. 
In the quasi-periodic case instead 
the r.h.s of \pref{hh1} can be arbitrarily small due to Umklapp (the momentum is defined modulo $2\pi$
for the presence of the lattice)
, so that terms with large  $n$ in \pref{hh1}
persist at any Renormalization Group iteration.  It is remarkable that dangerous processes in the infrared behavior are generated by the exchange of large momenta, which is a sort of ultraviolet-infrared mixing problem.
We will see in next section  that the relevance or irrelevance of such terms depends in a subtle way from number theoretical properties of the frequency $\o$ and the velocity of decay of the Fourier transform of the quasi-periodic potential.

\section{Irrelevance of the non resonant terms}

According to the above analysis, it is natural to distinguish 
in the effective potential \pref{ep} between two kind of terms; in the {\it resonant terms}
there is conservation of the momentum measured from the Fermi points, that is
the r.h.s. of \pref{hh1}  is vanishing
$
\sum_i\r_i p_F+
2 n \pi \o+2l \pi=0$; 
when the above condition is violated the terms are non resonant. We show now that if $\o$ is irrational
the non resonant terms are irrelevant, even if dimensionally relevant or marginal.
Roughly speaking the reason is that, by the Diophantine condition \pref{d},   the r.h.s. of \pref{hh1}
 is very small only if $n$ is very large; this can produce a gain factor, provided that the decay of the harmonics of the potential is fast enough.

In order to put on a quantitative basis the above idea
it is convenient, 
given a Feynman graph, 
to consider a maximally connected subset of lines corresponding 
to propagators 
with scale $h\ge h_v$ with at least a scale $h_v$, and we call it {\it cluster} $v$; 
the $n^e_v$ lines external to the cluster $v$ have scale smaller then $h_v$. 
Given a non maximal cluster $v$ with scale $h_v$, there is surely a cluster $v'$ with scale $h_{v'}>h_v$
containing it.
The clusters are therefore subgraphs in which the propagators carries a scale larger
than the external lines, that is the momentum measured from the Fermi points in the internal lines
is larger than in the external; they are a standard tool in renormalization theory to avoid the so called overlapping divergences. We call $\bar m_v$ is the number of $u,\n$ vertices  internal to the cluster $v$ and not of any smaller one;
in the cluster $v$ there are $k_{2,v}$
vertices $u e^{i \s_i \pi\o x}$, $\s_i=\pm$ such that $N_v=\sum_i \s_i$, so that $|N_v|\le k_{2,v}$.
To each Feynman graph is associated a hierarchy of 
clusters;
inside each cluster 
$v$ there are $S_v$ maximal clusters, that is clusters   
contained only in the cluster $v$ and not in any smaller one, or trivial clusters given by a single vertex. Each of such inner clusters are connected 
by a tree of propagators with scale $h_v$; by integrating the propagators in the tree and bounding the others, and using that ,
$\int d\xx |g^{(h)}_\r(\xx)|\le C\g^{-h}$ and that $|g^{(h)}_\r(\xx)|\le C\g^{h}$ we get 
that each graph of order $k$ contributing to $W^{(h)}_{m,n}$ is bounded by the sum over the scales of
\be C^k U^{k_1} u^{k_2} \g^{(2-m/2)h}
\prod_v \g^{(h_v-h_{v'})D_v
}\prod_v \g^{-h_v\bar m_v}\label{paz}\ee
where $k$ is the perturbative order, $k=k_1+k_2$ and $D_v=2-n^e_v/2$ is the {\it scaling dimension} of the cluster $v$.
The estimate \pref{paz} (a versions of Weinberg theorem for this model)
says that no infrared divergence are present
in the thermodynamic limit {\it provided that} there are no inner subgraphs $v$ with four or two external lines and there are only quartic interactions. Indeed if 
$D_v<0$ and $\bar m_v=0$ for any $v$ one can sum over the scales $h_v$ obtaining a finite result:
on the contrary if $D_v=0$ for some $v$ one gets a factor $|h|$ corresponding to a logarithmic divergence
and if $D_v=-1$ a factor $\g^{-h}$ summing over the scales.

In getting the above bound we have however not taken into account that $\o$ is irrational and verifies 
the Diophantine condition. Indeed
if $v$ is a non resonant cluster with $m$ external lines and verifying 
$\sum_{i=1}^{m}\e_i \r_i p_F
+2 N_v \pi \o+2 l\pi\not=0$ then, by using \pref{d}
\bea
&&m\g^{h_{v'}}\ge ||\sum_{i=1}^m\r_i k'_i||\ge\nn\\ 
&&||\sum_{i=1}^m \r_i n_F\pi\o+
2 N_v \pi\o ||\ge C_0(|n_F|+|N_v|)^{-\t}
\label{ll}\eea
which implies that
\be
|N_v|\ge C \g^{-h_{v'}\over \t}\label{xcx}
\ee
On the other hand $N_v=\sum_i \s_i$ and $|N_v|\le k_{2,v}$ so that in a non resonant cluster
there are necessarily 
a large number of $u$ vertices $k_{2,v}\ge \tilde C \g^{-h_{v'}/\t}$, and the associated factor $u^{k_{2,v}}$
is therefore very small. By \pref{xcx} we get, for $c<1$
\be c^{-k_2}\le \prod_v c^{-k_{2,v} 2^{h_{v'}-1}}\le \prod_v c^{-C 2^{h_{v}}\g^{-h_{v}/\t}
S^{NR}_v}\label{ccc}\ee
where $S^{NR}_v$ are the non resonant clusters contained in $v$.
Therefore for each non-resonant cluster $v$ with 2 or 4 external lines we have, by
choosing $\g^{1\over \t}/2=\g^\h>1$, than such factor is bounded by $\prod_v  \g^{4 h_v S^{NR}_v}$;
this is consequence of the fact $e^{-|\log c |C 2^{-\h  h}}\le {\g^{N h}\over [\log c C]^N]e^N}$. 
We can use the factor $\prod_{v,n^e_v>4} \g^{(h_v-h_{v'})D_v }$
to sum over the scales $h_v\le h_{v'}$; regarding the cluster with two or four external lines we distinguish between resonant clusters $v\in R$ and non resonant clusters $v\in NR$; in the second case
\be
\prod_{v\in NR, n^e_v=2,4} \g^{(h_v-h_{v'})D_v }\prod_v  \g^{2 h_v S^{NR}_v}\le
\prod_{v\in NR,n^e_v=2,4} \g^{h_v} 
\ee
so that the sum over the scales of the non resonan clusters with 2 or 4 external lines can be done. 
Finally the resonant clusters $v\in R$ is the
union of $R_1$ and $R_2$, with $R_1$  containing at least one non-resonant vertex and $R_2$ containing only  resonant vertices; therefore
\bea
&&\prod_{v\in R,n^e_v=2,4} \g^{(h_v-h_{v'})D_v }\prod_v  \g^{2 h_v S^{NR}_v}\label{fon}\le\\
&&[\prod_{v\in R_1,n^e_v=2,4} \g^{(h_v-h_{v'})D_v }\g^{h_v}][\prod_{v\in R_2,n^e_v=2,4} \g^{(h_v-h_{v'})D_v }]\nn
\eea
Therefore the only source of infraed divergences is
due a) to the resonant clusters with $n^e_v=2,4$ containing only resonant clusters; b)
to the resonant clusters with $n^e_v=2$ containing at least one non-resonant cluster.
The factor
$c^{k_2}$ is absorbed in $C^k$.
We see that there are no infrared divergences provided that there are no resonant subgraphs
with 2 or 4 external lines; in other words all the non resonant terms are irrelevant. 

A similar argument could be repeated even if all the harmonics are present in the
quasi periodic potential, that is by choosing $\phi_x=\sum_n \hat\phi_n e^{2i n\pi \o x}$, provided that $\hat\phi_n$ decays exponentially fast, for instance 
$|\hat\phi_n|\le e^{-\x |n|}$, see \cite{M1}. On the other hand, some kind of fast decaying condition is necessary for proving the irrelevance of the non resonant terms; if the decay is too slow 
the above argument does not provide any 
gain for the non resonant terms.
This is the case of
Fibonacci potential considered in \cite{G2},\cite{G3}, in which the Fourier coefficients decay only 
as $O(n^{-1})$.

\section{Renormalized expansion}

In the previous section we have identified the dangerous terms producing infrared divergences
in thermodynamic limit; such divergences has to be removed in order to get physical informations from the expansions. 
We have to set-up a different integration procedure in which the  
resonant terms which are dimensionally relevant or marginal are {\it renormalized}; in this way one produces an expansion in terms of running coupling constants in which no infrared divergences are present.
Of particular importance are the quadratic resonant terms with $\r_1=-\r_2$, corresponding to the generation of a gap.
Note that only when $n_F=1$ the initial interaction $V$ contains such terms, 
but if $n_F>1$ they are generated by higher order terms in the RG iterations.
It is convenient then to add to the effective action a term of the form
\be
\sum_\r\int d\kk \s(\kk)\psi^+_{\kk,\r}\psi^-_{\kk',-\r}-\sum_\r\int d\kk \a(\kk)\psi^+_{\kk,\r}
\psi^-_{\kk',-\r}\ee
and include the first term in the free integration so that the propagator becomes massive; 
$\a$ is chosen so that the flow of the resonant quadratic terms is bounded.
At the end we impose the  condition $\a(u,U,\s)=\s$ determining
$\s(u,U)$, so proving the generation of the gap in the original problem.
In the case $n_F=1$ this is of course not necessary ($\a=\s=0$).

The integration procedure of the generating functions
starts, as before, integrating the ultraviolet field $\psi^{(u.v.)}$; 
we can include the $\s$ term 
in the free integration so that
the generating function has the form (we set $\phi=J=0$ and $n_F>1$)
\be e^{\WW(0,0)}=
\int P_{1,\s_0}(d\psi^{(\le 0)})e^{\VV^{(0)}(\psi^{(\le 0)})}
\ee
where
$P_{1,\s_0}(d\psi^{\le 0})$ which has propagator given by
\bea
&& g^{(\le 0)}_{\r,\r'}(\xx-\yy)= \int d\kk e^{-i \kk (\xx-\yy)}\chi_0(\kk)\times\\ 
&&\begin{pmatrix}
&-i
k_0+ v_F\sin k'+c(k') & \s_0(\kk')\\ &\s_0(\kk') & -i k_0- v_F \sin k' +c(k')
\end{pmatrix}^{-1}_{\r,\r'}\nn\label{prop1}
\eea
with $\s_0$ is equal to $\s$ (or to $u$ is $n_F=1$)up to higher order corrections, 
$c(k)=\cos p_F(\cos k'-1)$, and $\VV^{(0)}$ is given by an expression
similar to \pref{ep}, in which now the kernels
$W^{(0)}_{m,n}$ are series in $u,U,\a,\n$.
%
%
We describe the RG procedure iteratively. Assume that we have integrated 
$\psi^{(0)},\psi^{(-1)},...,\psi^{(h+1)}$ obtaining
\be
e^{\WW(0,0)}=
\int  P_{Z_h,\s_h}(d\psi^{(\le h)}) e^{\VV^{(h)}(\sqrt{Z_h}\psi^{\le h})}\label{ss}
\ee
where 
\bea
&& g^{(\le h)}_{\r,\r'}(\xx-\yy)= {1\over Z_h}\int d\kk e^{-i \kk (\xx-\yy)}\chi_h(\kk)\times\\ 
&&\begin{pmatrix}
&-i
k_0+ v_F\sin k'+c(k') & \s_h(\kk')\\ &\s_h(\kk') & -i k_0- v_F \sin k' +c(k')
\end{pmatrix}^{-1}_{\r,\r'}\nn\label{prop1}
\eea
and $\VV^{(h)}$ is a sum of monomials with kernels $W^{(h)}_n$ which are
%
expressed as 
sum of renormalized Feynman diagrams with vertices or associated to the running coupling constants 
$\l_k,\d_k,\n_k,\a_k$, $k>h$ or to the non resonant terms $u$ present in $\VV$; they depend also from $\s_k,Z_k$ through the propagators. The single scale propagator is equal to \pref{rel}
up to terms with the same scaling properties and an extra $\s_h\g^{-h}$ in the diagonal,
and an extra $(\s_h\g^{-h})^2$ in the non diagonal component.
We have to extract from 
the effective potential $\VV^{(h)}$ the non irrelevant part which is called $\LL \VV^{(h)}$; that is we write
$
\VV^{(h)}=\LL \VV^{(h)}+\RR \VV^{(h)}
$
with $\RR=1-\LL$. $\LL$ acts non trivially 
only on the resonant terms. In particular $\RR$ renormalize the non irrelevant subgraphs eliminating the infrared divergences.
The presence of a mass has the effect that we integrate up 
a mass scale $h^*$ defined by the condition $\g^{h^*}=\s_{h^*}$, $\s_h=\s_h(0) $;
we will see that $\s_h\sim \s_0 \g^{\h_\m h}$ with $\h_\m=O(\l)$.
The scales 
$\le h^*$ can be integrated in a single step.
 
When $m=2$, $\r_1=\r_2$
the localization operation is
\be
\LL W^{(h)}_{\r,\r}(\kk')=P_0 W^{(h)}_{\r,\r}(0)+\kk'\partial P_0 W^{(h)}_{\r,\r}(0) \label{nnnn}
\ee
where $P_0$ sets all the $\s_k=0$ in the propagators in the kernels; note that $P_0 W^{(h)}_{\r,\r}(\kk')$ minus the r.h.s. of 
\pref{nnnn} is ${1\over 2}\kk'^2\partial^2 P_0 W^{(h)}_{\r,\r}$, so that a gain 
$\g^{2(h_{v'}-h_v)}$ is produced by the $\RR$ operation, which is sufficient to make the 
corresponding sum over $h_v$ convergent.
On the other hand 
in $1-P_0$ there is an extra $({\s_{h_v}\over\g^{h_v}})^2$ (coming or from the correction to a diagonal propagator, or from a couple of diagonal propagators) producing an extra 
$[{\s_{h^*}\over\g^{h^*}}]^2 ({\s_{h_v}\over \s_{h^*}}{\g^{h^*}\over\g^{h_v}})^2
\le 
\g^{{3\over 2}(h^*-h_v)}$ which again  eliminate the divergence.
The term $P_0 W^{(h)}_{\r,\r}(0)$ contribute to the running coupling constant
$\n_h$,
%
the term $\partial_0 P_0 W^{(h)}_{\r,\r}(0)$ to the wave function renormalization $Z_h$
%
%
and $\partial_1 P_0 W^{(h)}_{\r,\r}(0)-\partial_0 P_0 W^{(h)}_{\r,\r}(0))$
to the Fermi velocity renormalization.
%
%
Regarding the bilinear terms with $\r_1=-\r_2$ 
%
we write 
$W^{(h)}=W_{a,\r,-\r}^{(h)}+W_{b,\r,-\r}^{(h)}$, where $W_{a,\r,-\r}^{(h)}$
contains only graphs with vertices $\l_k,\d_k,\n_k$ so that in $W_{b}^{(h)}$, the rest,
there is at least an $\a_k$ or an irrelevant vertex term $u$;
we include the linear part in $\s$ of $W^{(h)}_{a,\r,-\r}(0)$
(there is at least one $\s$ and the higher order terms
have an extra $(\s_h\g^{-h})^2$)
in the renormalization of $\s_h$ and 
$W^{(h)}_{b,\r,-\r}(\kk')$ in the renormalization of $\a_h$.
%
%
%
%
%
%
%

Finally for the kernels with $m=4$ and $\sum_i \r_i\e_i=0$ 
$\LL W^h_{4}(\underline\kk)=P_0 W^h_{4}(0)$ which is included in the renormalization of $\l_h$.
%
In conclusion with the above decomposition of the effective potential and after a redefinition of the effective wave function renormalization and gap , \pref{ss} is equal to
\be
\int  P_{Z_{h-1},\s_{h-1}}(d\psi^{\le h}) e^{-\bar\LL\VV^{h}(\sqrt{Z_{h-1}}\psi^{\le h})-\RR (\sqrt{Z_{h-1}}\psi^{\le h})}\label{ss1}
\ee
with 
\be
\bar\LL\VV^{h}=\g^h \n_h F_{\n}^h+\g^h\a_h F_\a^h+\d_h F_\d^h+\l_h F_\l
\ee
where 
\bea
&&
F_\l=
\int \prod_{i=1}^4 d\kk'_i \psi^+_{\kk'_1,+} \psi^-_{\kk'_2,+}
\psi^+_{\kk'_3,-} \psi^-_{\kk'_4,-}\d(\sum_i \s_i \kk'_i)\\
&&F_\n^h=\sum_\r \int d\kk' \psi^+_{\kk',\r} \psi^-_{\kk',\r}\quad F_\a^h=\sum_\r \int d\kk' \psi^+_{\kk',\r} \psi^-_{\kk',-\r}
\nn\\
&&F_\d^h=\sum_\r \int d\kk' \r v_F \sin k' \psi^+_{\kk',\r} \psi^-_{\kk',\r}
\eea
%
The terms $\int d\kk' \psi^+_{\kk',\r} \psi^+_{\kk',-\r}$
and $\int d\kk' (-ik_0+v_F\r\sin k')  \psi^+_{\kk',\r} \psi^+_{\kk',\r}$ have been included in the free integration to produce the renormalization of $\s_h$ and $Z_h$. In conclusion we write 
$P_{Z_{h-1},\s_{h-1}}(d\psi^{\le h-1})=
P_{Z_{h-1},\s_{h-1}}(d\psi^{\le h-1})P_{Z_{h-1},\s_{h-1}}(d\psi^{(h)})$ and we can integrate the field 
$\psi^{(h)}$ obtaining an expression similar to  
\pref{ss} from which the procedure can be iterated. 
A similar analysis can be repeated in presence of the source term $\int d\kk'_1 d\kk'_2
W_{2,1}(\kk'_1,\kk'_2) J \psi^+_{\r_1,\kk'_1}\psi^+_{\r_1,\kk'_2}$ ; in that case
\be
\LL W^{(h)}_{2,1}(\kk'_1,\kk'_2)= P_0 W^{(h)}_{2,1}(0,0) 
\ee
which is included in the non oscillating or oscillating renormalization of the density $Z^{(1)}_h$ or $Z^+_h$ depending if $\r_1=\r_2$ or $\r_1=-\r_2$.

The outcome of the above procedure is that the correlations can be expressed by renormalized diagrams functions  of the running coupling constants and such that the renormalization $\RR$
acts on the resonant clusters eliminating the infrared divergences; indeed 
the sum over the scales can be done and on gets an estimate for graphs with $l$ vertices
$ C^l\e_h^{k}
\g^{(2-m/2)h}$
where $\e_h=max_{k\ge h}(|\l_k|,|\a_k|,|\n_k|,|\d_k|)$ and $\hat z_v=3/2$; therefore the sum over the scale give a finite result.
Note that, from the above estimate, the graph with an irrelevant term have an extra $\g^h$.
The above bound has been derived for a single graph but, due to cancellations related to the anticommutativity of fermions, 
it is also valid for the sum of all graphs of order $n$; it is then suitable to establish analyticity, 
provided that $\e_h^{k}$ is bounded in $h$; in order to verify this we have to study the flow of the running coupling constants.

\section{The flow of the running coupling constants}

The expansion described in the previous section is convergent provided that the effective couplings are small.
The flow of the relevant couplings $\a_h,\n_h$ is controlled by choosing properly $\a,\n$ and one gets 
\be
\a_h(\kk')\sim u^{2 n_F}\g^h\quad \n_h\sim (u+|U|)\g^h
\label{ppp1}
\ee
and that $\l_h,\d_h$ remain close to their initial value $\l_0,\d_0$. 
To prove \pref{ppp1} we note that the flow equation for 
$\n_h$ is 
\be
\n_{h-1}=\g^{-h}(\n_0+\sum_{k=0}^h \g^k \b^{(k)}_\n)
\ee
and by choosing $\n_0$ so that $\n_0+\sum_{k=0}^{-\io} \g^k \b^{(k)}_\n=0$ one gets 
\be
\n_{h-1}=-\g^{-h}\sum_{k=-\io}^h \g^k \b^{(k)}_\n\label{ppp}
\ee
To $\b^{(k)}_\n$ contribute: a)terms with at least an irrelevant vertex, which are $\sim \g^k$; b)terms depending only from the running coupling constants $\l,\d$; in this case the contributions
containing only propagators $g_L^{(k)}$ gives a vanishing contributions, in the others there is an extra $\g^h$; c)terms containg at least $\a_h,\n_h$ which are $\sim\g^h$.
Using \pref{ppp} we get the second of \pref{ppp1}.
Similarly we write 
\be
\a_{h-1}(\kk')=\g^{-h}(\a_0(\kk')+\sum_{k=0}^h \g^k \b^{(k)}_\a(\kk'))
\ee
and by choosing 
$(\a_0-\sum_k \g^h W^{(h)}_{b,\o,-\o}(\kk'))=0$ we get 
\be
\a_{h-1}(\kk')=-\g^{-h}\sum_{k=-\io}^h \g^k \b^{(k)}_\a(\kk')\ee
By construction in  $\b^{(k)}_\a$ there is at least an irrelevant term or an $\a_k$ so that $\a_h\sim u^{2 n_F}\g^h$.
The lowest order contribution  to $\int d\kk W^h(\kk')\psi^-_{\kk',+}\psi^-_{\kk',-}$
is obtained by the chain graph with $2 n_F$
$u$ vertices propagators carrying momentum $k'+(n_F-1)\pi\o,....,k'-(n_F+1)\pi\o$, and the corresponding contribution to $\a_0$, obtained setting $\kk'=0$, is 
$a_{n_F} u^{2n_F}$ with, $n_F>1$
\be
a_{n_F}(\kk')= \prod_{k=1}^{2n_F}{1\over -i k_0+\cos(\pi\o(n_F-k))-\cos n_F\pi\o} 
\ee
which is non vanishing.  Note that we can renormalize simply subtracting $a_{n_F}(0)$, as is sum of clusters containing at last
a non resonant cluster, so by \pref{fon} a gain $\g^{h_{v'}-h_v}$ is sufficient.
Regarding the higher order terms the are at least $O(u^{2n_F+1})$ or $O(U u^{2 n_F})$; by imposing $\s_0=\a_0$ we get 
\be
\s_0(\kk')=u^{2n_F}(a_{n_F}(0)+u F_1(u)+U F_2(U,u))
\ee
with $F_1, F_2$ bounded; therefore the term $a_{n_F}$ dominates if $U,u$ are small enough. 
Regarding the flow for $\l_h,\d_h$ again we decompose the beta function in a part depending only from 
$\l_h,\d_h$ and propagators $g_{L}^{(k)}$, see \pref{rel}, and a rest which is $\sim \g^k$ (as there is a
$r^{(h)}$ \pref{rel} or a $\a_h,\n_h$, or irrelevant terms).
This second part is summable while the first coincides with the Luttinger model one and is asymptotically vanishing so that
\be
\l_h\to _{h\to-\io}\l_{-\io}(U)=U (\hat v(0)-\hat v(2p_F))+...\label{fff}
\ee
and similarly $\d_h\to _{\to-\io}\d_{-\io}(U)$.
The flow equation for $\s$ 
\be
{\s_{h-1}(\kk')\over\s_h(\kk')}=1+\b^{h}_{\s}(\kk)\equiv e^{\tilde \b^h_\s(\kk')}
\ee
implies
\be
{\s_{h-1}(\kk')\over\s_0(\kk')}=e^{\sum_{k=h}^0 \tilde \b^k_\s(\kk')}
\ee
We  
divide the beta function for $\s_h$ in a part $\b^h_{\a,a}$ depending from 
 $g_{L}^{(k)}$, see \pref{rel}, and containing only the vertex $\l_k,\d_k$ and a rest $\b^h_{\a,b}$ which is $\sim \g^h$ 
(as there is a
$r^{(h)}$ \pref{rel} or a $\n_h$, or irrelevant terms); moreover  $\tilde \b^k_\s(\kk')-\tilde \b^k_\s(0)\sim \g^{h-k}$ as
$\kk'\sim \g^h$. Therefore
\be
\s_h\sim \s_0(\kk') \g^{\h_\m h}\quad\quad \h_\m=1-{\l_{-\io}\over 2\pi v_F}+...
\ee
where 
the exponent $\h_\m$ takes contribution only from $g_L$ depends only from $\l_{-\io},\d_{-\io}$.
Similarly the wave function and density renormalizations behaves as 
\be
Z_h\sim \g^{\h_z h}\quad Z^{(1)}_h\sim \g^{\h_z h}\quad Z^{(+)}_h\sim \g^{\h_+ h}
\ee
where $\h_z,\h_+$ 
takes no contribution contribution only from $g_L$ and $\l_{-\io},\d_{-\io}$; again the contribution with $\a>k,\n_k$ or from $r^{(h)}$ \pref{rel} are $O(\g^h)$.

\section{Gap renormalization and scaling relations}

The Renormalization Group is iterated up to a scale $h^*$ such 
that $\s_{h^*}\sim \g^{h^*}$; as $\s_h=\s_0 \g^{\h_\m h}$ then 
$\g^{h^*}=(\s_0)^{1\over 1-\h_\m}$.
All the scales $<h^{*}$ can be integrated in a single step, as the scaling properties of the propagator $g^{<h^*}(\xx)$ are the same as $g^{(k)}(\xx)$ for $k\ge h^*$. 
As a consequence, the 2-point propagator 
decays faster that any power with rate 
$\g^{h^*}=(\s_0)^{1\over 1-\h_\m}$; this provides an estimate of the gap 
\be
\D_{n,U}\sim (\D_{n,0}+F_{n,U})^{X_n}\label{b}
\ee
with $X_n={1\over 1-\h_\m}$, $\D_{n,U}$ is the gap in the non interacting case and 
$F_{n,u,U}$ is of order $U u^{2 n_F}$. In conclusion when $p_F=n\pi\o$ the 2-point function decays  for large distances as \be
|<\psi^-_\xx\psi^+_\yy>|
\le {1\over |\xx-\yy|^{1+\h_z}}{C_N\over (\D_{n,U}|\xx-\yy|)^N}\ee
for any integer $N$. Similarly the density-density correlations can be written as
\be
<\r_\xx\r_\yy>=G_1(\xx,\yy)+\sin p_F(x-y) G_2(\xx,\yy)+G_3(\xx,\yy)
\ee
where 
\be
|G_2(\xx,\yy)|\le 
{1\over |\xx-\yy|^{2 K_n}}{1\over (\D_{n,U}|\xx-\yy|)^N}\ee
with $2K_n=2(1+\h_+-\h_z)$; $G_1$ and $G_3$ verify similar bounds with $K_n$ replaced by $1$
and $3/2$ respectively.
By construction 
$
\h_\m=\h_+-\h_z$ and using that $X_n={1\over 1-\h_\m}$ 
we get finally
\be
X_{n}={1\over 2-K_{n}}\label{hhh}
\ee
Note that, for $u,U$ small
\be
\D_{n,U}/\D_{n,0}\sim u^{-2n(1-K_{n})\over 2-K_{n}}\ee
and $K_{n}>1$ for attractive interactions and $<1$ for repulsive interactions; therefore 
the relative size of the gaps is greatly decreased or increased depending from the attractive or repulsive nature of the interaction. Finally,
as the exponents depends from $\l_{-\io},\d_{-\io}$ and $g_L^{(h
)}$ only, they verify in addition the usual Luttinger liquid relations, whose validity can be established by Ward Identities \cite{BFM}; in particular the exponent in the 2-point function verifies $\h_z={2-K_{n}-K_{n}^{-1}\over 2}$.

\section{Conclusions}

A non perturbative Renormalization Group analysis of the interacting Aubry-Andr\'e model
for fermions, as realized for instance in cold atoms experiments \cite{I2}, has been performed.
We prove that, for weak coupling,
all the infinitely many gaps persist, and no insulator-metal transition is present even if the chemical potential is in correspondence of small gaps of the non interacting spectrum. As only a single harmonic is present in the Aubry-Andr\'e potential, 
the gaps in the non 
interacting spectrum can be generated at very high order and are therefore very small;
the relative size of small gaps is therefore greatly enhanced or depressed 
depending from the nature of the interaction, an effect which could be seen in experiments.
The gaps are renormalized by critical exponents 
verifying exact scaling relations, and this is
in apparent agreement with numerical simulations in \cite{S}.
The behavior of the Aubry-Andr\'e model, in which all the gaps persists,
is therefore rather different from the scenario proposed in  \cite{G3} in the case of Fibonacci potential, in which even an arbitrarily weak interaction closes small gaps and produces an insulating-metal transition.  It is likely that an extension of the methods adopted here can 
be also used for a full analysis of the Fibonacci case beyond lowest order truncation, to check the validity of this scenario.

%
%

\bibliographystyle{amsalpha}

\end{document}